\documentstyle[epsfig,aps]{revtex} 
\begin{document}
\twocolumn[\hsize\textwidth\columnwidth\hsize
           \csname @twocolumnfalse\endcsname]
\title{Testing Dirac-Brueckner models in collective flow of 
heavy-ion collisions}
\author{
T. Gaitanos$^{1}$, C. Fuchs$^{2}$, 
H.H. Wolter$^{1}$, and Amand Faessler$^{2}$
}                   
\address{
$^{1}$ Sektion Physik der Universit\"at M\"unchen, 
Am Coulombwall 1, D-85748 Garching, Germany\\
$^{2}$Institut f\"ur Theoretische Physik der Universit\"at T\"ubingen, 
Auf der Morgenstelle 14, D-72076 T\"ubingen, Germany
}
\maketitle
\begin{abstract}
We investigate differential in-plane and out-of-plane flow observables 
in heavy ion reactions at intermediate energies from $0.2\div 2$ AGeV 
within the framework of relativistic BUU transport calculations. The 
mean field is based on microscopic Dirac-Brueckner-Hartree-Fock (DBHF) 
calculations. We apply two different sets of DBHF predictions, those 
of ter Haar and Malfliet and more recent ones from the T\"ubingen group, 
which are similar in general but differ in details. The 
latter DBHF calculations exclude spurious contributions 
from the negative energy sector to the mean field which 
results in a slightly softer equation of state and a 
less repulsive momentum dependence of the nucleon-nucleus 
potential at high densities and high momenta. For the application 
to heavy ion collisions in both cases non-equilibrium features 
of the phase space are taken into account on the level of the 
effective interaction. The systematic comparison to experimental 
data favours the less repulsive and softer model. 
Relative to non-relativistic approaches 
one obtains larger values of the effective nucleon mass. This 
produces a sufficient amount of repulsion to describe 
the differential flow data reasonably well.
\end{abstract}
\section{Introduction}
\label{intro}
Relativistic heavy-ion collisions have been extensively investigated to 
determine the nuclear 
equation-of-state (EOS) far away from saturation and at finite temperature, 
using semi-classical transport models of the Boltzmann-type \cite{buu,rbuu}. 
The nuclear EOS which enters into a transport description via density and 
momentum dependent mean fields has mostly been based on 
phenomenological considerations by adjusting the 
parameters to nuclear matter saturation properties 
and to the momentum dependence of the 
empirical nucleon-nucleus optical potential \cite{hama}. 
Such parametrisations, as e.g. Skyrme-type potentials \cite{skyrme},  
imply different extrapolations to high and low 
densities and high momenta, 
which should be tested in heavy ion reactions. Thus there have 
been many efforts to determine the density and momentum dependence 
of the nuclear mean field by studying the different aspects of 
the collective flow in intermediate energy heavy ion collisions between 
$0.1 -2$ AGeV (SIS-energies) 
\cite{skyrme,ritter97,hermann,mbc94,zheng99,nemeth99,giessen1,giessen2,dani00,larionov00,gait96,gait99}.

On the other hand, it is well known that hadrons generally change their 
properties in the medium. This basic feature is already incorporated in the 
simplest version of a relativistic hadronic model for nuclear matter, namely linear 
Quantum Hadron Dynamics (QHD) \cite{hs86}, where the effective nucleon mass 
drops with density. To obtain a reasonable compressibility scalar 
self-interaction terms are introduced and finite nuclei are well described \cite{ring96}. 
At much higher densities, i.e. for the description of neutron stars, 
also non-linear terms of the vector mean field are required \cite{stars}. 
These different treatments reflect the inherent uncertainties in density 
extrapolations away from the saturation point. More recent and systematic 
approaches try to fix the relevant terms, e.g. by density functional expansions
of generalised QHD lagrangiens \cite{IJMPE,FST} which effectively 
incorporate the basic features of chiral symmetry and its breaking. 
However, predictions for high 
densities remain questionable since effective field theory provides a low 
density expansion scheme valid in the  
vicinity of the nuclear saturation point and below \cite{FST,lutz00}.

An alternative approach to the density and momentum dependence of the mean field 
is provided by 
microscopic many-body models. Here the nucleon-nucleon (NN) interaction is fixed 
by free NN-scattering and no parameters are adjusted to the nuclear matter problem. 
In the relativistic Dirac-Brueckner-Hartree-Fock (DB) approach 
\cite{DBHF,btm90,harmal,tueb2,tueb,DDH} the NN-interaction is based on 
modern one-boson-exchange (OBE) potentials \cite{bonn} 
and the in-medium ladder 
diagrams are summed self-consistently. This 
approach describes the nuclear matter saturation properties very 
reasonably, albeit not perfectly. There exist good arguments why DBHF should 
be still reliable at higher densities. As pointed out in \cite{machleidt99} 
the nonlocal structure of the OBE potentials accounts already effectively 
on the two-body level for some features of many-nucleon terms, e.g. 
intermediate $\Delta$-excitations, and higher order effects cancel to 
large extent. Then the success of non-linear QHD lagrangiens can be understood in 
that these models parametrise phenomenologically the density dependence 
of the microscopic DB predictions \cite{fuchs96}. However, the constraints from 
finite nuclei on the explicit form of the different fields are limited since 
the single particle potential results from the 
cancellation of large scalar and vector potentials. Only the spin-orbit 
interaction allows to constrain the magnitude of the effective 
mass \cite{ring96,IJMPE,DDH}. On the other hand, in energetic heavy ion reactions 
scalar and vector fields are decoupled by their different 
Lorentz transformation properties, in the sense that they can be tested 
independently, and additional information on the structure of the potential 
can be obtained. Thus within the DB framework there is the chance to attempt 
a {\it unified} description of different nuclear systems, i.e. free NN-scattering, 
nuclear matter, finite nuclei and heavy ion collision. 

However, relativistic Brueckner calculations are not straightforward and the 
approaches of various groups \cite{btm90,harmal,tueb2,tueb} 
are similar but differ in detail, 
depending on solution techniques and the particular approximations made.  
In the present work we therefore want to study the DB predictions 
for the nuclear mean field in more detail in heavy ion collisions. 
In previous studies \cite{gait96,gait99} a qualitative agreement with 
collective flow observables has been found. 
Recently more detailed experiments have been performed for 
differential components of the collective flow \cite{exp00}. 
In particular, the rapidity and 
transverse momentum dependence of collective flow has attracted 
much theoretical 
interest because of their strong sensitivity to the momentum dependence of the 
nuclear mean field \cite{dani00}. Here we test the self energies, 
i.e. the nuclear EOS, from two different DB calculations, from ter Haar and Malfliet 
\cite{harmal} and from a more recent study performed by the 
T\"ubingen group \cite{tueb2,tueb}. 

In heavy ion collisions a further difficulty arises 
due to the non-equilibrium features of the phase space configurations. 
This has been discussed extensively in ref. \cite{gait96,gait99}. 
In a fully consistent treatment, one would have to solve the coupled set of DB 
equations for the effective interaction in non-equilibrium nuclear matter 
configurations simultaneously with the kinetic equations 
for the evolution of a phase space distribution \cite{btm90}. 
However, such a procedure has not been realized yet, and further approximations 
are necessary in heavy ion collisions. 
In the local density approximation (LDA) the nuclear matter 
mean fields are directly used in the transport calculation. 
However, this approximation is not reliable enough at intermediate energies, 
because the local momentum space is highly anisotropic during a large part of the 
heavy ion collision \cite{gait99,essler,puri92}. 
A better approximation is the colliding nuclear matter (CNM) approach, where 
the phase space anisotropies are parametrised by two inter-penetrating 
nuclear matter currents, i.e. the local momentum space is given by two Fermi-spheres, 
or covariant Fermi-ellipsoids, with given Fermi momenta and 
a relative velocity \cite{sehn}. 
In Ref. \cite{sehn} a method was developed to extrapolate nuclear matter DB results 
to CNM configurations. The CNM self energies are applied to heavy ion collisions 
in the Local (phase space) Configuration Approximation (LCA) 
\cite{gait96,gait99,gmat2}, where the 
anisotropic phase space is locally parametrised by a CNM configuration. 
In this paper the LCA approximation is only briefly discussed, 
details can be found in \cite{gait96,gait99,sehn}. 
We shall discuss the density and momentum dependence of the DB mean fields and 
their application to the CNM approximation in terms of an effective equation of state 
\cite{fuchs01}. 

This paper is organised as follows: 
The DB formalism and the different DB models are reviewed in sec. II. The CNM 
approximation is discussed in sec. III. In  section \ref{sec:1} the transport equation 
and their numerical realization is outlined. Then we discuss the 
density and momentum dependence of the DB mean fields in ground state 
nuclear matter, and the 
LCA approximation for anisotropic phase space configurations in heavy ion collisions. 
Section \ref{sec:3} contains the results of transport calculations based on the DB 
mean fields of Refs. \cite{harmal,tueb} in the LCA approximation. 
We compare different components of collective flow 
with respect to its energy, centrality, transverse momentum and rapidity dependence. 
It is found that both models are able to qualitatively describe the experimental 
results, but in details one finds model dependences on the collective flow observables.

\section{The DB approach}
Brueckner theory provides a microscopic model which accounts for 
two-body correlations in the ladder approximation in medium in 
the Bethe-Goldstone or the Bethe-Salpeter equation in the  relativistic case, 
\begin{equation}
{\cal T} = V+ i\int  V GG{\cal T} 
\label{DB1}
\quad .
\end{equation}
The correlations of the Green functions, or wave functions respectively,  
are shifted to the effective in-medium interaction, i.e. the 
${\cal T}$-matrix (or ${\cal G}$-matrix) \cite{btm90}. The in-medium 
propagator obeys a Dyson equation
\begin{equation}
G = G^{0} + \int G^{0} \Sigma G 
\label{DB2}
\end{equation}
and is dressed by a self-energy $\Sigma$ obtained in Hartree-Fock 
approximation from the ${\cal T}$-matrix 
\begin{equation}
\Sigma = -i\int {\cal T}G 
\label{DB3}
\quad .
\end{equation}
The coupled set of equations (\ref{DB1}-\ref{DB3}) has to be 
solved self-consistently. In this procedure the bare interaction $V$, 
iterated in the Bethe-Salpeter equation, is 
sandwiched between dressed in-medium 
spinors. This feature is absent in non-relativistic approaches 
and introduces an additional density dependence 
which is responsible for the significantly improved saturations 
properties compared to non-relativistic ${\cal G}$-matrix 
calculations \cite{gmat1,baldo89}. The non-relativistic Brueckner 
approach leads to too large saturation densities (e.g. $\rho_{\rm sat} = 
0.2~{\rm fm}^{-3}$ in \cite{gmat1} and $\rho_{\rm sat} = 
0.24~{\rm fm}^{-3}$ in \cite{baldo89}) and predicts 
a rather small compressibility (K=180 MeV in \cite{gmat2,gmat1}). 
The introduction of $3$-body-forces can in principle improve on this in 
the non-relativistic case \cite{baldo89}. 

In contrast to the phenomenological approaches like QHD \cite{hs86} 
and to effective field theory \cite{IJMPE,lutz00} this approach is essentially 
parameter free. The only free parameters are those of the 
NN-interaction, namely those of the realistic OBE potentials, which are fixed by 
the free scattering problem \cite{bonn}. The success of the DB model 
indicates that it already accounts for the most important set of 
diagrams to describe nuclear matter. A further inclusion of particle-hole 
correlations around the DB mean field also can ensure thermodynamical 
consistency in the form of the Hugenholtz-van-Hove theorem \cite{zuo99,lenske96}. 
  
In the present work we employ the results of two different DB calculations, 
those of Ref. \cite{harmal} and the more recent ones of Ref. \cite{tueb}, denoted 
as DBHM and DBT, respectively, in the following. 
We will briefly characterise the difference between these two calculations. 
One difference is the use of different OBE potentials; in \cite{harmal} 
a version of the Groningen potential and in \cite{tueb} 
the Bonn A potential. In both cases the same set of six non-strange mesons 
with masses below 1 GeV is used and the fit to the NN phase shifts is of 
similar quality, however, the actual model parameters (coupling constants and 
form factors) are different. The main difference between these two approaches 
has a more complicated origin, which is discussed in detail 
in \cite{tueb2,tueb}. 
The DB structure equations (\ref{DB1}-\ref{DB3}) are matrix equations 
in spinor space. To determine the Dirac structure of the self energy, i.e. 
the scalar ($\Sigma_{s}$) and a vector ($\Sigma^{\mu}$) contributions, 
\begin{equation}
\Sigma_{\alpha\beta} = \mbox{ {\small 1}\hspace{-0.3em}1}_{\alpha\beta}\Sigma_{s}-
\gamma^{\mu}_{\alpha\beta} \Sigma_{\mu}
\label{decom}
\end{equation}
the ${\cal T}$-matrix has to be decomposed into its Lorentz components, i.e. 
scalar, vector, tensor, etc. contributions. This procedure is not free 
from ambiguities. Due to identical matrix elements for positive energy states 
pseudo-scalar and pseudo-vector 
components cannot uniquely be disentangled for the on-shell ${\cal T}$-matrix. 
However, with a pseudo-scalar vertex the pion couples maximally to negative energy 
states which are not included in the standard Brueckner approach. This 
is inconsistent with the potentials used since the OBE potentials are based 
on the no-sea approximation. Hence, pseudo-scalar contributions 
due to the one-$\pi$ exchange lead to large and 
spurious contributions from negative energy states. In 
\cite{tueb2} it was shown that such spurious contributions 
dominate the momentum dependence of the nuclear 
self-energy, and, in particular, lead to an artificially strong momentum dependence 
inside the Fermi sea. It was further demonstrated in \cite{tueb2} 
that the method used in \cite{harmal} fails to 
cure this problem and in \cite{tueb} a 
new and reliable method was proposed to remove those spurious contributions 
from the ${\cal T}$-matrix. 
\begin{table}
\begin{center}
\caption{
Saturation properties of nuclear matter, i.e. 
Fermi-momentum $k_F$, saturation density $\rho_{sat}$, 
binding energy per particle $E/A$, effective mass $m^*$ and the compression modulus
K in the DBHF calculations of \protect\cite{harmal} (DBHM) and  \protect\cite{tueb} (DBT).
}
\begin{tabular}{cccccc}
      & $k_F$       & $\rho_{sat}$  &$E/A$    &  $ m^*$ &   K  \\
      & [fm$^{-1}$] & [fm$^{-3}$]  &[MeV]  & [MeV]   & [MeV] \\
\hline
 DBHM      & 1.343 & 0.164 & -13.6  &  558    & 250 \\
 DBT       & 1.39  & 0.185 & -16.1  &  637    & 230 \\
\end{tabular}
\label{table2}
\end{center}
\end{table}
The saturation properties of the two DB calculations are given in Table 1. 
It is seen there that the results of ter Haar and Malfliet (DBHM) \cite{harmal} 
give a slightly better saturation density compared to \cite{tueb} (DBT) but 
too little binding. DBT, in contrast, gives a good binding energy and also meets 
the empirical range of saturation. 
The stiffness of the EOS expressed by the compression moduli is similar 
for both approaches. A significant difference 
can be observed for the magnitude of the effective mass but both values are 
consistent with the knowledge 
from finite nuclei on the strength of the spin-orbit force 
($ 500~{\rm MeV} \leq m^* \leq 700~{\rm MeV} $) \cite{FST}. 

We also note that the relativistic effective 
mass, i.e. the Dirac mass $m^* = M - \Sigma_s$ given in Table 1, 
should be distinguished from the effective mass 
$m^{*}_{\rm NR}$ which in non-relativistic approaches is used to 
classify the non-locality of the mean field \cite{dani00,larionov00}. 
The latter is defined by 
\begin{equation} 
m^{*}_{\rm NR} = |{\bf k}| 
\left( \frac{\partial k^0}{\partial |{\bf k}|} \right)^{-1}
\end{equation} 
and is approximately $m^{*}_{\rm NR} \approx k^{*}_0 = \sqrt{ {\bf k}^2 + m^{*2}}$.  
At $\rho_0$ the two models DBHM/DBT yield the following 
values $m^{*}_{\rm NR} (k_F) / M =  0.63 / 0.73$ 
which can be compared to the parameters used in refs.  
\cite{dani00,larionov00}.

\begin{figure}
\centering
\resizebox{0.45\textwidth}{!}{
  \includegraphics{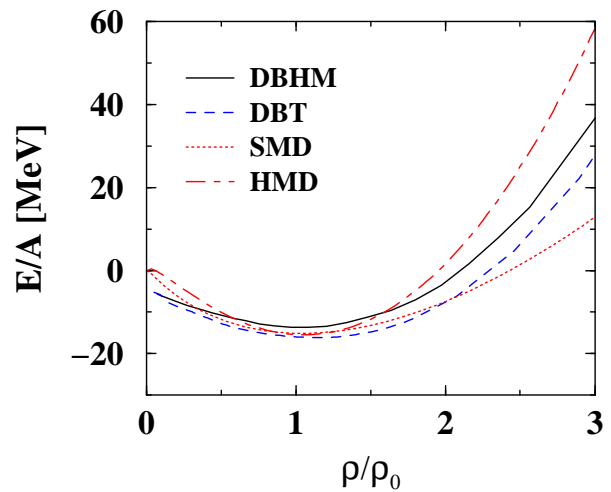}
}
\caption{Equation of states in the DB approaches. Solid: DB calculations from 
\protect\cite{harmal}, dashed: DB calculations from \protect\cite{tueb}. 
For comparison also the soft/hard (SMD/HMD) 
momentum dependent Skyrme parameterisations with a compression modulus of 
K=200/380 MeV \protect\cite{skyrme} are 
shown.}
\label{Fig8}   
\end{figure}

The corresponding equations-of-state are shown in Fig.\ref{Fig8}. 
Both models have a similar 
density dependence, however, due to the higher binding energy DBT 
lies generally below DBHM. At high densities 
the softer character of DBT becomes a little more 
pronounced. For comparison also the widely used Skyrme 
parameterisations (soft/hard, denoted by SMD/HMD) with 
compression moduli of K=200/380 MeV \protect\cite{skyrme} are shown. 
It is seen that in particular DBT is close to the soft Skyrme 
parameterisation up to about $2\rho_0$ where it starts to 
become more repulsive.

\begin{figure}
\centering
\resizebox{0.45\textwidth}{!}{
  \includegraphics{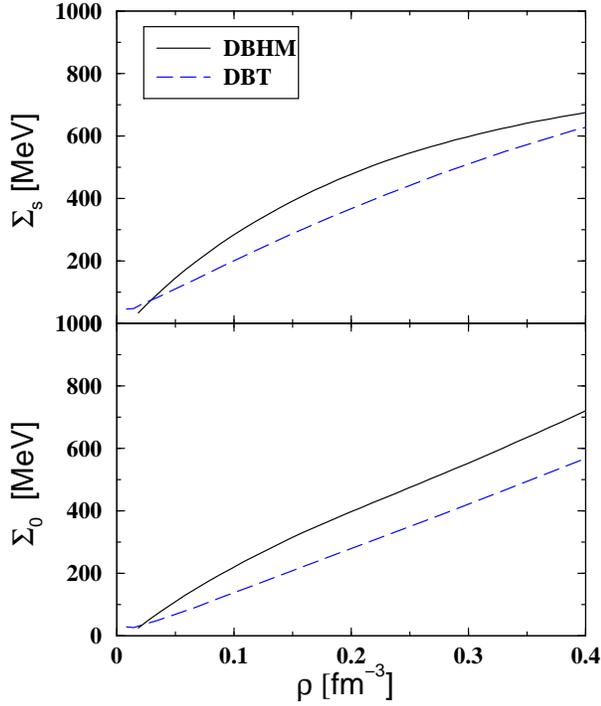}
}
\caption{Scalar (top) and vector (bottom) self energies. The solid lines 
are DB calculations from \protect\cite{harmal}, the dashed lines 
from \protect\cite{tueb}. 
}
\label{Fig16}   
\end{figure}
\begin{figure}
\centering
\resizebox{0.45\textwidth}{!}{
  \includegraphics{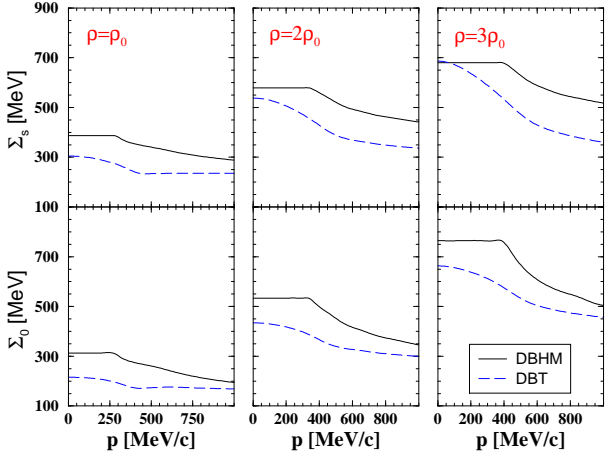}
}
\caption{Momentum dependence of the scalar (top) and vector (bottom) 
self energies at $\rho=1/2/3 \rho_0$. 
The solid lines are DB calculations from \protect\cite{harmal} 
and the dashed curves from 
\protect\cite{tueb}.
}
\label{Fig6}   
\end{figure}
Fig.\ref{Fig16} shows the density dependence of the mean field. 
In the DBT approach the fields are generally smaller by about 
50-100 MeV compared to DBHM. The same trend can be seen from 
Fig.\ref{Fig6} where the momentum dependence of the scalar and 
vector potentials at densities $\rho_0=0.16~{\rm fm}^{-3}$, 
$2\rho_0$ and $3\rho_0$ 
is shown. In both calculations scalar and vector fields 
decrease with increasing momentum in a similar way. 
Generally, the explicit momentum dependence is moderate at 
densities $\rho \leq \rho_0$ but becomes pronounced at 
higher densities. In ref. \cite{harmal} constant values of $\Sigma$ 
are taken for momenta below the Fermi momentum whereas the results   
of \cite{tueb} reflect the full momentum dependence outside and 
inside the Fermi sea. The difference in the magnitude of the 
fields  $\Sigma_s$ and  $\Sigma_0$ in the two approaches can be 
traced back to the different projection schemes discussed above. 
With the correct and complete 
pseudo-vector description for the pion contributions the fields 
are dominated by the $\sigma$ and $\omega$ contributions and the other 
mesons $\pi$(pseudo-vector coupling),$\rho ,~\eta ,~\delta$ give 
only small corrections \cite{tueb2,tueb}. 

\begin{figure}
\centering
\resizebox{0.45\textwidth}{!}{
  \includegraphics{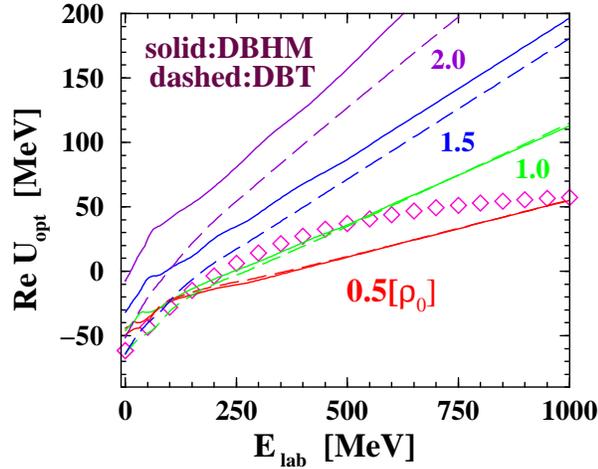}
}
\caption{Energy dependence of the optical potential. The solid lines 
are DB calculations from \protect\cite{harmal}, the dashed lines 
from \protect\cite{tueb}. 
}
\label{Fig9}   
\end{figure}

Fig.\ref{Fig9} shows the real part of the optical Schroedinger-equivalent 
nucleon potential, defined as 
\begin{equation}
U_{\rm opt} 
=  - \Sigma_s  + \frac{k^0}{M} \Sigma_{0} 
        + \frac{\Sigma_s^2  - \Sigma_{\mu}^2}{2M}~,
\label{uopt}
\end{equation}
in nuclear matter as a function 
of its laboratory energy $E_{\rm lab} = k^0 - M$. Note that 
the definition of an optical potential is not unique in the 
literature, e.g. in refs. \cite{dani00,li93} an optical 
potential is defined as the difference of the single-particle energies in the medium 
and in free space $U= k^0 - \sqrt{M^2 +{\bf k}^2}$. 
The optical potential defined by 
Eq. (\ref{uopt}) is the Schroedinger-equivalent relativistic 
potential \cite{harmal} and can be covariantly defined by 
$U_{\rm opt}  = (k^{2}_\mu - M^2)/2M 
= ( (k^{*}_\mu +\Sigma_{\mu})^2 - M^2)/2M$ as a Lorentz scalar. Even a 
momentum independent vector potential $\Sigma_{0}$ (as in the mean field 
approximation of QHD) 
leads to a linear energy dependence of the optical potential  (\ref{uopt}), 
i.e. a momentum dependence $\frac{\Sigma_0 (\rho)}{2M^2}p^2$. 
The explicit momentum dependence of the 
DB fields falls asymptotically as $\Sigma_{0,S} \sim (A + B/p)$ 
which still leads to a linear increase of $U_{\rm opt}$ 
at large energies. As seen in Fig.~\ref{Fig9} the DB model reproduces 
the empirical optical potential \cite{hama} extracted from 
proton-nucleus scattering for nuclear matter at $\rho_0$ reasonably well 
up to a laboratory energy of about 0.6-0.8 GeV. 
However, it is seen that the saturation behaviour at large momenta cannot 
be reproduced by DB calculations. In heavy ion 
reactions at incident energies above 1 AGeV such a saturation behaviour is 
required to reproduce transverse flow observables \cite{giessen2}. 
Thus, DB mean fields start to become unrealistic around 
1 AGeV. There exists presently no microscopic relativistic calculation 
which is able to reproduce this saturation behaviour of the optical 
potential. Therefore we restrict our investigation of transverse flow 
observables to an energy range where the DB fields can be safely applied. 
At higher energies one has to rely on phenomenological approaches where 
the strength of the vector potential is artificially suppressed, e.g. by 
the introduction of additional form factors \cite{giessen2}. 

One should be aware that the empirical optical potential involves densities 
around $\rho_{0}$ and does not 
completely constrain the mean fields that enter in a heavy ion collision, 
which involve large values of momentum and density. 
As a common feature relativistic 
DB calculation show a strong and repulsive momentum dependence 
also at high densities \cite{harmal,tueb,li93} whereas, e.g. the 
non-relativistic ${\cal G}$-matrix \cite{baldo89} has 
a much less repulsive high density behaviour. 

\begin{figure}
\centering
\resizebox{0.45\textwidth}{!}{
  \includegraphics{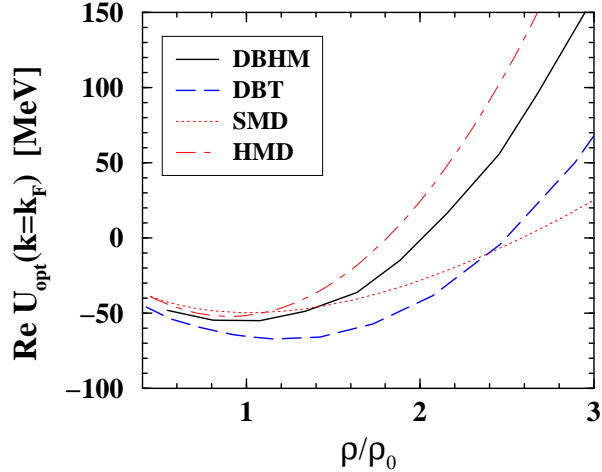}
}
\caption{Optical potential $U_{\rm opt} (\rho,k=k_F)$ 
at the Fermi surface as a function of density. The solid lines 
are DB calculations from \protect\cite{harmal}, the dashed lines 
from \protect\cite{tueb}. For comparison also the 
soft/hard (SMD/HMD) momentum dependent Skyrme parameterisations 
\protect\cite{skyrme} are shown.}
\label{Fig9b}   
\end{figure}
In first order the  
strength of the repulsion in the relativistic case is 
determined by the magnitude of the vector field. 
As can be seen from Fig.\ref{Fig9} the two 
approaches DBHM and DBT yield similar results at moderate densities 
$\rho \leq \rho_0$ but differences become substantial at high densities. 
The generally smaller fields of DBT result in a less repulsive 
potential at high densities. This behaviour becomes even more clear 
from Fig.\ref{Fig9b} where the optical potentials 
taken at the Fermi surface, i.e. at $k=k_F$, are shown as a function 
of the density. For comparison 
also the potentials of the soft/hard momentum dependent Skyrme 
parameterisations (SMD/HMD) \cite{skyrme} are given. 
For DBHM the interplay between density and momentum dependence of 
the optical potential is similar to the hard Skyrme force 
although the EOS is less repulsive at high densities. 
On the other hand DBT is even more attractive than the soft Skyrme force 
at low densities and becomes more repulsive than 
the soft Skyrme force only above $2\rho_0$. Hence the character of the 
mean field obtained by the two DB model calculations show essential 
differences although the corresponding EOSs are similar. 
The test of DB fields in heavy ion collisions
where high densities ($\rho \approx 2-3 \rho_0$) 
and momenta greater than the Fermi-momentum can be reached, 
should allow to differentiate between the two models. 

\section{The CNM approximation}
\begin{figure}
\begin{center}
\leavevmode
\epsfxsize = 8cm
\epsffile[70 280 550 490]{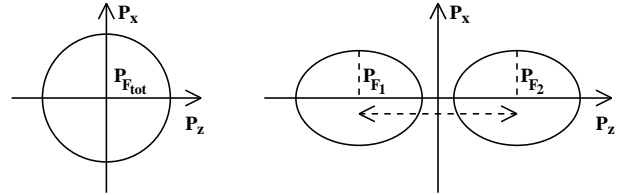}
\end{center}
\caption{
Schematic representation of the LDA (nuclear matter) and LCA 
(colliding nuclear matter) approximations.
}
\label{Fig0}
\end{figure}

The DB approach discussed in the previous section 
describes equilibrated nuclear matter which is 
characterised by one Fermi-sphere of a given Fermi-momentum. 
In a local density approximation (LDA) these self energies 
are directly inserted into the drift 
term of the RBUU equation. However, as discussed before, 
local momentum space anisotropies are a characteristic feature 
of energetic heavy ion reactions. The time scales where such anisotropies 
occur are comparable with the compression phase of the 
process \cite{gait99,essler,puri92}. It was found 
that the local anisotropic momentum space can well 
be parametrised by two inter-penetrating nuclear matter currents, i.e. 
by two Fermi-spheres in momentum space \cite{essler,puri92}, or the 
Colliding Nuclear Matter (CNM) \cite{sehn} which is 
schematically illustrated in Fig.\ref{Fig0}. The application 
of the relativistic DB model to CNM configurations 
has, however, not been realized yet, but only non-relativistic 
${\cal G}$-matrix calculations have been performed for CNM \cite{gmat1}. 
Therefore in Ref. \cite{sehn} a method was developed to extrapolate 
DB results to such CNM configurations. The CNM momentum distribution 
is constructed as a superposition of the single currents, i.e. 
\begin{eqnarray}
f_{12}({\bf k}^{*}) & = & f_{1}({\bf k}^{*}) + f_{2}({\bf k}^{*}) + \delta f 
\nonumber\\
                    & = & \Theta(k_{F_{1}}-k^{*}_{\mu}u_{1}^{\mu}) + 
                          \Theta(k_{F_{2}}-k^{*}_{\mu}u_{2}^{\mu}) + \delta f 
\label{f_cnm}
\quad ,
\end{eqnarray}
 where $f_{i}$ are the covariant momentum space distributions of the two NM currents 
and $\delta f=-f_{1}f_{2}$ is a correction term which takes into account Pauli-blocking 
effects in the overlap region of the two nuclear matter currents. 
The corresponding CNM self energies 
$\Sigma_{s,0}(k; \chi)$ depend explicitely on momentum and the configuration parameters 
$\chi \equiv \{k_{F_{1}},k_{F_{2}}, v_{rel} \}$ 
(Fermi-momenta $k_{F_{1,2}}$ and the relative velocity $v_{rel}$). 
Averaging the CNM configuration, Eq. (\ref{f_cnm}), over momentum leads to mean 
self energies which depend only on the parameters 
of the CNM momentum space distributions. 

\begin{figure}
\centering
\resizebox{0.45\textwidth}{!}{
  \includegraphics{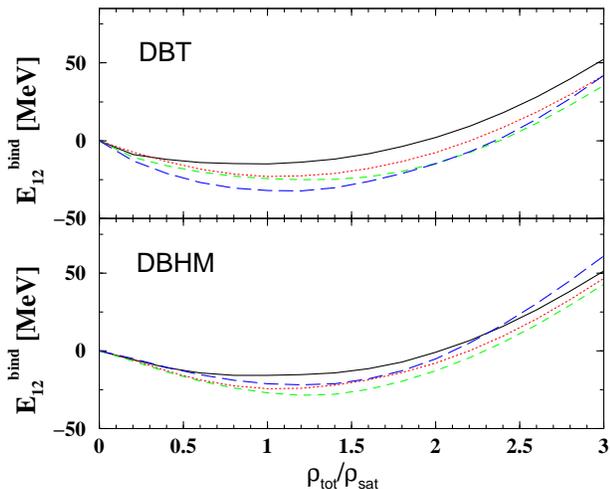}
}
\caption{Subtracted  EOSs for the CNM approximation using the DB results of Ref. 
\protect\cite{harmal} (top) and \protect\cite{tueb} (bottom). The Different 
curves give the CNM EOS's for different relative velocities 
($v_{rel}=0,~0.2,~0.4,~0.6$ for solid, 
dotted, dashed and long-dashed curves, respectively).}
\label{Fig15}   
\end{figure}

The consideration of anisotropy effects 
in the CNM approximation leads to non-equilibrium mean fields, 
which essentially differ from those of the equilibrium case. 
In analogy to the non-equilibrium self energies we construct a 
non-equilibrium EOS \cite{sehn}, i.e. 
an equation of state which depends on the CNM parameters. 
In order to compare the compression energies it is useful to 
subtract the  kinetic energy of the 
relative motion of the two nuclear matter currents 
which yields the ``subtracted'' binding energy 
$E_{12}^{bind}$ of the system. Fig.\ref{Fig15} shows these  
EOS's for different symmetric 
($k_{F} \equiv k_{F_{1}}=k_{F_{2}}$) CNM configurations. It is seen 
in both models that the 
effective non-equilibrium EOS is softer compared to the equilibrium 
EOS (solid curves for vanishing relative velocity). 
However, the different density and momentum dependence of the 
two DB models leads to a different magnitude of this softening effect, 
in particular with increasing relative velocity. 

This softening of the effective EOS in colliding nuclear 
matter can be understood by considering that in 
the participant zone of the reaction particles which belong to 
projectile and target are separated in momentum space. This 
geometrical effect works in a similar way as an additional degree 
of freedom and leads to a softening of the effective 
EOS experienced by the nucleons in such configurations. 
This type of phase space effect is not included in standard transport 
calculations for heavy ion collisions, 
even when momentum dependent interactions are 
used \cite{dani00,larionov00}. Phenomenological mean fields 
$ U({\bf x},{\bf k} ) = U_{\rm loc}(\varrho({\bf x}) ) + U_{\rm nonloc} ({\bf x},{\bf k})$
are usually composed by a local, density dependent potential 
$ U_{\rm loc}(\varrho({\bf x}) )$ and 
a non-local momentum dependent part 
$U_{\rm nonloc} ({\bf x},{\bf k}) = \int d^3 k' f({\bf x},{\bf k}') 
V({\bf x},{\bf k}-{\bf k}')$ 
with $V$ an effective momentum dependent two-body-interaction. Whereas 
$U_{\rm nonloc}({\bf x},{\bf k})$ accounts properly for the actual momentum 
space configurations $f({\bf x},{\bf k}')$, 
the local part does not depend on the momentum space. 
Consequently, $U_{\rm loc}(\varrho )$ reflects a density 
dependence which is correct in equilibrated nuclear matter but does 
not apply to anisotropic momentum space configurations. 
\nopagebreak
\section{The transport model}
\label{sec:1}
In the present work heavy ion collisions are treated by the relativistic 
(R)BUU equation \cite{rbuu,gait96,gait99} 
\begin{equation}
\left\{
   k_{\mu}^{*}\partial_{x}^{\mu} + 
   \left[ 
      k_{\nu}^{*}F^{\mu\nu}+m^{*}(\partial^{\mu}_{x}m^{*})
   \right]
   \partial^{k^{*}}_{\mu}
\right\}
f(x,k^{*}) = I_{coll}
\label{rbuu}
\quad .         
\end{equation}
This equation describes the evolution of a classical one-body phase space distribution 
$f(x,k^{*})$ under the influence of a self-consistent mean field, or the scalar and 
vector self energies $\Sigma_{s}$ and $\Sigma^{\mu}$. 
The self-energies determine effective momenta and masses of the 
dressed quasi-particles in the nuclear medium 
$k^{*\mu}=k^{\mu}-\Sigma^{\mu}$, $m^{*}=M-\Sigma_{s}$. 
The field strength tensor of the vector field 
$F^{\mu\nu}=\partial^{\mu}\Sigma^{\nu}-\partial^{\nu}\Sigma^{\mu}$ gives rise 
to a Lorentz force as in electrodynamics. 
This introduces in a most natural way a first order 
momentum dependence which in non-relativistic 
treatments has to be parameterised explicitely \cite{buu,skyrme,giessen1}. 
The collision term describes $2$-body collisions and is treated by 
cascade-like Monte-Carlo simulations, as in relativistic versions of 
the QMD-model \cite{rqmd}. We include 
the relevant nucleonic excitations at SIS energies, 
i.e. the $\Delta (1232)$ and $N^{*}(1440)$ resonances and their decay to one- and 
two-pion states. The cross-sections for elastic and inelastic scattering as well 
as differential cross sections are taken from Ref. \cite{Hu94} which 
are used also in QMD calculations at SIS energies 
\cite{hartnack98,uma97}. The drift term of the RBUU equation (\ref{rbuu}) is 
numerically treated in the 
relativistic Landau-Vlasov method (RLV)  \cite{fuchs95}. 
This is a test particle method, where the 
test particles are represented by manifestly covariant Gaussians in phase space.  
In order to reduce numerical fluctuations a number of $50-100$ test particles 
per nucleon was found to be sufficient here. 
The nuclei are initialised to fit density profiles 
obtained from self-consistent Thomas-Fermi calculations \cite{fuchs95}. For 
both models (DBT and DBHM) we adjust the testparticle distributions 
to the same reference distribution. The corresponding initialisations 
are stable over the durations of the considered reactions. 
Energy-momentum conservation is fulfilled with an accuracy 
of $3-5$ \% of the initial kinetic centre-of-mass 
energy of the colliding nuclei. 

The anisotropic phase space effects discussed above are incorporated 
in heavy ion collisions by applying the CNM or non-equilibrium DB mean fields 
in the framework of a Local Configuration Approximation (LCA) 
\cite{gait96,gait99,puri92}. In this approach the 
phase space is parametrised locally by a CNM configuration where the invariant 
configuration parameters $\chi$ are directly determined from the phase 
space distribution $f(x,k^{*})$. Transport calculations have shown 
that the collective flow is reduced if the non-equilibrium effects 
are taken into account \cite{gait99} which is consistent 
with a softening of the effective EOS in heavy ion collisions as discussed above. 

\section{Collective flow effects}
\label{sec:3}
In this section different types of collective flow observables are investigated 
in transport calculations using the DBHM/DBT mean fields in the 
local phase space configuration approximation (LCA). 
To compare with experiments the same methods 
are used for centrality selection and reaction plane determination, 
and the theoretical results are subjected to filter routines to 
simulate the experimental detector efficiencies, if necessary. Since we mainly 
compare to results from the FOPI Collaboration \cite{hermann,exp00,fopi95,andronic99} 
these correspond to the FOPI (Phase-I or Phase-II) set up. These filters are 
sensitive to fragment distributions. Thus we also generate fragments 
in the final state of the  reaction 
(at $\approx 100-200$ fm/c depending on the incident energy) using a 
phenomenological phase space coalescence model. 
The coalescence parameters in coordinate and 
momentum space are separately adjusted for both models DBT and DBHM 
to fit the experimental mass distributions of light fragments 
($Z\leq 3$) \cite{gait99,gait00}. After the cluster formation we 
apply the same procedure for the reaction plane reconstruction as 
the FOPI group. The corresponding corrections are very close to those obtained 
with the IQMD model \cite{crochet2}.
\begin{figure}
\centering
\resizebox{0.45\textwidth}{!}{
  \includegraphics{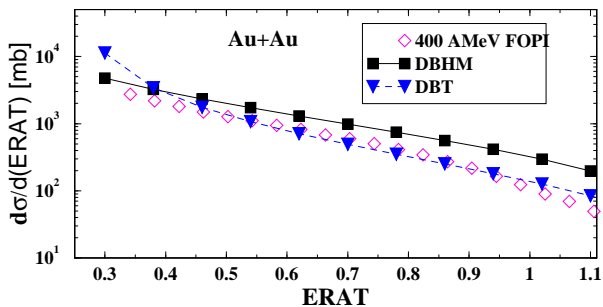}
}
\caption{$ERAT$ cross sections for protons in 
Au+Au reactions at 0.4 AGeV beam energy. 
Transport calculations with the DB mean fields from Refs. 
\protect\cite{harmal} (DBHM) 
\protect\cite{tueb} (DBT) are compared with the FOPI data \protect\cite{fopi95}.}
\label{Fig13}   
\end{figure}
Criteria for the determination of the centrality class of an 
event are the multiplicity of charged particles 
(PMUL) and/or the ratio of transversal to 
longitudinal energy ($ERAT$) \cite{reisdorf97}. As an example, in Figs. \ref{Fig13} 
and \ref{Fig14} these observables are shown for protons at an 
energy of 0.4 AGeV and compared to FOPI data \cite{fopi95,reisdorf97}. 
Fig. \ref{Fig13} displays the differential cross 
section $d\sigma/d(ERAT)$. Large $ERAT$ values correspond 
to central reactions whereas small values indicate 
semi-peripheral and peripheral reactions \cite{gait99,fopi95,reisdorf97}. 
A qualitative agreement with 
experiment is achieved for both mean fields, with DBHM somewhat 
overestimating the data. At small ERAT values below about 0.2 
the quantity $d\sigma /dERAT$ is 
strongly affected by trigger effects and drops very rapidly. 
As discussed in \cite{fopi95} model calculations tend 
to strongly overestimate the data below 0.3, which is due to 
detector cuts which remove the projectile remnant. Since 
such remnants, i.e. very heavy clusters, are not formed in 
transport calculations the cross section is overpredicted in 
very peripheral collisions although the total reaction 
cross section coincides with the experimental one. 
\begin{figure}
\centering
\resizebox{0.45\textwidth}{!}{
  \includegraphics{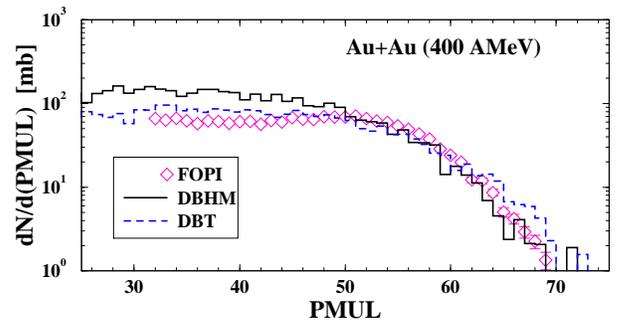}
}
\caption{Multiplicity of charged particles. The solid lines 
are DB calculations from \protect\cite{harmal}, the dashed lines 
from \protect\cite{tueb}.
}
\label{Fig14}   
\end{figure}
In Fig. \ref{Fig14} the multiplicity distributions 
$d\sigma/d(PMUL)$ are shown. Also here the agreement with experiment 
is reasonable for DBHM and quite good for DBT. Particularly, we can 
reproduce the plateau (in logarithmic scale) of these 
distributions  which is used in the centrality selection. 
Altogether, the good agreement with experiment makes the 
centrality selection reliable using either the PMUL or 
$ERAT$ observable. In the the following we will 
mainly use the $PMUL$ criterion. The correlation 
between multiplicity and centrality intervals is defined as in experiments 
in the following way: the lower limit of the highest multiplicity 
bin, called $PM5$, is fixed at half of the plateau value, and the 
remaining multiplicity range is divided into four equally spaced intervals, 
denoted by $PM4$ to $PM1$. $PM5$ then corresponds to most central reactions, 
and $PM4$ and $PM3$  to semi-central and peripheral ones, 
respectively \cite{reisdorf97}.

\subsection{Nuclear stopping}

We start the flow analysis with the longitudinal distributions 
which are characterised in terms of the rapidity 
$Y_{cm}=\frac{1}{2}\ln{(1+\beta_{{\rm c.m.}})/(1-\beta_{{\rm c.m.}})}$. 
Here the normalised rapidity 
$Y^{(0)} = Y_{{\rm c.m.}}/Y_{{\rm c.m.}}^{{\rm proj}}$ is 
considered. 
\begin{figure}
\centering
\resizebox{0.45\textwidth}{!}{
  \includegraphics{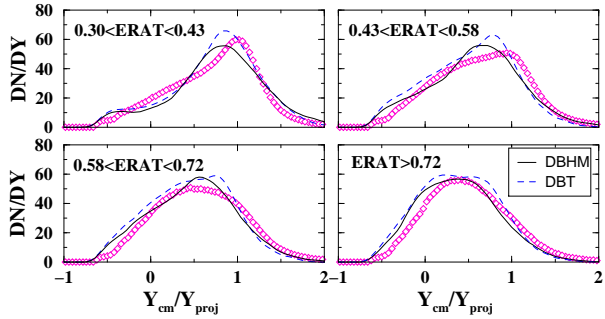}
}
\caption{Rapidity distributions for Au+Au collisions at $0.4$ AGeV beam energy. The 
different centrality intervals run from  peripheral (upper left) 
to central (lower right) collisions. The solid lines 
are DB calculations from \protect\cite{harmal}, the dashed lines 
from \protect\cite{tueb}.
}
\label{Fig7}   
\end{figure}
The rapidity distributions are shown in Fig.\ref{Fig7} for different centrality 
classes (using the $ERAT$ selection \cite{fopi95}). 
This observable is strongly affected by detector cuts 
\cite{fopi95,reisdorf97}, which is reflected in the asymmetry 
of these distributions relative to the c.m.-rapidity, 
due to the angular limitations of the FOPI-detector (Phase-I) 
\cite{reisdorf97}.

It is seen that both models are able to generally reproduce the 
experimental results. Stopping is influenced most strongly by 
the choice of the NN cross sections, which are the same in both models. 
An indirect influence also 
from the EOS can be expected  since a softer and less repulsive EOS leads to more 
compression and thus to more collisions. However, the differences are not very 
pronounced which reflects the fact that the EOSs are 
still similar in the explored density regime and do not differ too 
much even at the maximal densities reach in central reactions 
($\rho \approx 2-3~\rho_0$). In peripheral collisions, 
where the deflection of spectator matter by the repulsive 
momentum dependent component of the mean field plays a more 
dominant role, the rapidity distributions show bigger differences. 
Similar trends have been observed in in Ref. \cite{fopi95} with 
soft/hard Skyrme forces. 

\subsection{In-plane flow}
\begin{figure}
\centering
\resizebox{0.45\textwidth}{!}{
  \includegraphics{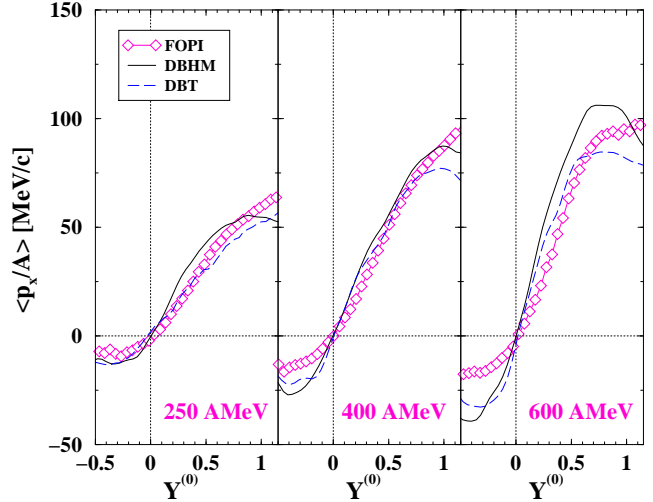}
}
\caption{Mean in-plane transverse flow of protons versus normalised rapidity 
for semi-central (PM4) Au+Au collisions at 0.25, 0.4 and 0.6 
AGeV beam energy. The solid lines 
are DB calculations from \protect\cite{harmal}, the dashed lines 
from \protect\cite{tueb}. The experimental results 
are taken from \protect\cite{crochet}.}
\label{Fig5}   
\end{figure}

\begin{figure}
\centering
\resizebox{0.45\textwidth}{!}{
  \includegraphics{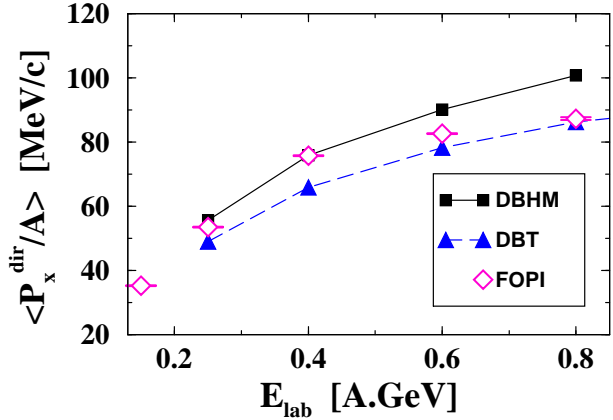}
}
\caption{Excitation function 
of the mean directed in-plane flow for semi-central (PM4) Au+Au 
collisions (data from \protect\cite{crochet}). The solid lines 
are DB calculations from \protect\cite{harmal}, the dashed lines 
from \protect\cite{tueb}.
}
\label{Fig10}   
\end{figure}

Next we consider the emission of matter projected onto the 
reaction plane described by the mean in-plane or 
sideward flow \cite{ritter97,hermann}. 
Fig. \ref{Fig5} compares the mean in-plane proton flow $<p_x /A>$ 
in semi-central (PM4) Au+Au 
collisions at incident energies of 0.25, 0.4 and 0.6 AGeV 
to the FOPI data from \cite{crochet}. Both models reproduce the energy 
dependent increase of the proton flow and are generally in good 
agreement with the data. We observe that DBHM 
leads to a stronger in-plane flow in the spectator rapidity region 
whereas the slope near mid-rapidity $Y^{(0)} \sim 0$ is less 
affected by the differences in density and 
momentum dependence of the mean fields. At 0.6 AGeV 
DBHM starts to overpredict the slope of the in-plane flow 
whereas DBT is still in reasonable agreement with experiment. 
This reflects again the more repulsive character of the DBHM forces 
which becomes more pronounced 
with increasing energy, in particular in the spectator 
region $Y^{(0)} \sim 1$. On the other hand, DBT describes the in-plane flow 
well around midrapidity but slightly under predicts it in the spectator 
region. 
The situation is more clearly seen when the 
mean directed flow $P_{x}^{dir}$, i.e. the 
in-plane flow integrated over the forward hemisphere ($Y^{(0)} \ge 0$), 
is considered. In Fig.\ref{Fig10} the excitation function of 
$P_{x}^{dir}$ is compared to the FOPI data \cite{crochet}, 
again for PM4. $P_{x}^{dir}$ is a 
measure for the overall repulsion experienced by the reaction. 
The comparison with data shows that both microscopic models can explain the 
experimental results relatively well. However, above 0.4 AGeV the 
experimental excitation function of $P_{x}^{dir}$ shows a 
saturation behaviour which is not completely reproduced. 
Here, as already seen in Fig.\ref{Fig5}, DBT is in good agreement 
with experiment at higher energies but underpredicts the flow at 
lower energies. For DBHM the situation is just opposite. 
Due to the FOPI acceptance $P_{x}^{dir}$ is dominated 
by the bounce-off of spectator matter at $Y^{(0)} \sim 1$ 
(see the corresponding rapidity distributions in Fig.\ref{Fig7}) which explains 
the low value of DBT at 0.4 AGeV. The more repulsive character of the 
DBHM mean fields, in particular at high densities, 
produces a bounce-off of the spectator remnants in the reaction 
plane, resulting in too high transverse momenta near spectator 
rapidities at energies above 0.6 AGeV. Since the maximal densities 
reached in the model calculations are changing very little 
between 0.4 and 0.8 GeV the slope of the $P_{x}^{dir}$ excitation 
function is determined mostly by the momentum 
dependence of the mean field and less 
by the density dependence of the EOS. This may be considered as a 
way to test the average momentum dependence of the models. 
However, as discussd below an integrated observable 
reflects the average dynamics which is dominated by densities 
around $\rho_0$. To extract more decisive information on the 
momentum dependence at supranormal densities differential 
observables should be considered. 
As a differential observable the dependence of the in-plane flow on transverse 
momentum $p_{t}$ has recently attracted great interest. While 
the global behavior of the transverse flow, as expressed by 
$P_{x}^{dir}$ or $<p_{x}/A>$, gives an average over the entire evolution 
of the collision, the $p_t$ dependence allows to obtain 
information on different stages of the reaction. High $p_t$ nucleons, 
but also pions \cite{uma97}, originate from the 
early and high density phase of the reaction. This is a general 
feature of heavy ion collisions and seems to hold at 
bombarding energy from SIS \cite{larionov00} to AGS and 
SPS energies \cite{zabrodin00}. In Ref. \cite{dani00} is was 
pointed out that the $p_t$ dependence of the transverse flow 
in peripheral reactions is particularly sensitive to the momentum 
dependence of the nuclear mean field at supranormal densities. 
Consistent with this observation, in \cite{larionov00} it was 
demonstrated that the emission time of high $p_t$ particles at 
midrapidity coincides well with the high density phase of the 
reactions. To be more quantitative, in Fig. 13 we consider the 
correlation between the averaged density $<\rho_B>$ at which 
the particles are emitted and the transverse momentum$p_t$. Exemplarily 
a semi-peripheral (b=6 fm) Au+Au reaction at $0.4$ AGeV is considered. 
Since particles interact with the surrounding matter over the 
entire duration of the reaction more or less strongly 
- with the mean field and by binary collisions - it 
is {\it a priori} not clear how to define an emission time 
and, correspondingly, a density from 
which the particles carry information. It is, however, natural to 
select that time and that corresponding density where the particles 
experience their most violent changes in momentum. This can be done by the 
following quantity  
\begin{equation}
< \rho_B > = \sum_{i} \int dt \rho_B ({\bf x_i},t) 
\frac{|{\dot {\bf p}_i}|}{|{\bf p}_i|}~~ / ~~
\sum_{i} \int dt \frac{|{\dot {\bf p}_i}|}{|{\bf p}_i |}~~,
\label{ptrho}
\end{equation}
where the index $i$ runs over all baryons. 
Eq.(\ref{ptrho}) samples over the densities weighted by the 
relative changes of the momenta. This is a generalisation of the 
``freeze-out'' density because it takes the whole 
reaction history into account. Since experimentally 
only the final state momenta are detected, this quantity is useful 
to establish a correlation between final observables 
and the stages of the reaction on which the particles carry information.
The resulting emission densities are generally moderate because 
one averages over the entire phase space, 
i.e. over participant and spectator regions, and over the complete reaction 
history including the dilute expansion phase. 
Applying different $p_t$-cuts ($p_{t}^{(0)}=p_{t}/p_{c.m.}^{proj}$) 
one sees that the low $p_t$-particles and hence the bulk of the 
nucleons experiences a dynamical evolution which takes mainly 
place at densities below saturation density when integrated over the 
complete space-time of the reaction. The high $p_t$-particles 
($p_{t}^{(0)}>0,5$),  in contrast, are governed 
by larger densities. As a consequence, they probe the momentum 
dependence of the optical potential at supranormal densities.
In Fig. 13 as in the calcualtions below (Fig.\ref{Fig1}) we 
considered projectile rapidities, but the observed behaviour 
is even more pronounced in the midrapidity region. 
\begin{figure}
\centering
\resizebox{0.45\textwidth}{!}{
  \includegraphics{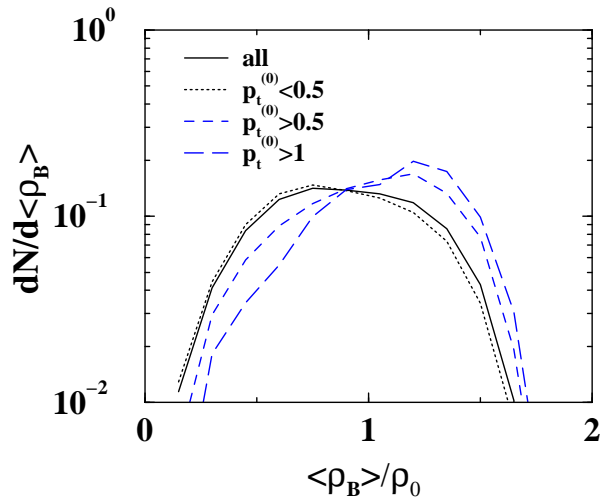}
}
\caption{The density distribution which is correlated with the dynamical 
history of the particles is shown for different $p_{t}^{(0)}$ cuts for 
a semi-central (b=6 fm) Au+Au collisions at $0.4$ AGeV.
}
\end{figure}
The flow and its $p_{t}^{(0)}$-dependence has been discussed in 
terms of a Fourier analysis of the 
experimental azimuthal distribution \cite{povo98}
\begin{equation}
\frac{dN}{d\phi}(p_{t}^{(0)},~Y^{(0)}) = v_{0}(1 + 2v_{1}cos(\phi)+2v_{2}cos(2\phi))
\label{fit}
\quad .
\end{equation}
In Eq. (\ref{fit}) $v_{0}$ is a normalisation constant, 
$v_{1}(p_{t}^{(0)},~Y^{(0)})$ describes the 
in-plane collective flow and $v_{2}(p_{t}^{(0)},~Y^{(0)})$ the emission 
perpendicular to the reaction plane, also called elliptic flow. 
The quantities $v_{1,2}$ can be determined directly 
as $v_{1}=<p_{x}/p_{t}>$ and $v_{2}=<(p_{x}^{2}-p_{y}^{2})/p_{t}^{2}>$, where 
$p_{t}=\sqrt{p_{x}^{2}+p_{y}^{2}}$ is the transverse momentum per nucleon. 

Fig. \ref{Fig1} shows the $p_{t}^{(0)}$-dependence of the 
in-plane flow ($v_{1}$) at normalised rapidities 
$0.5 \leq Y^{(0)}\leq 0.7$ for semi-central (PM4) Au+Au reactions at 
$0.4$ AGeV incident energy for protons ($Z=1$) and for 
light fragments ($Z=2+3$). 
\begin{figure}
\centering
\resizebox{0.45\textwidth}{!}{
  \includegraphics{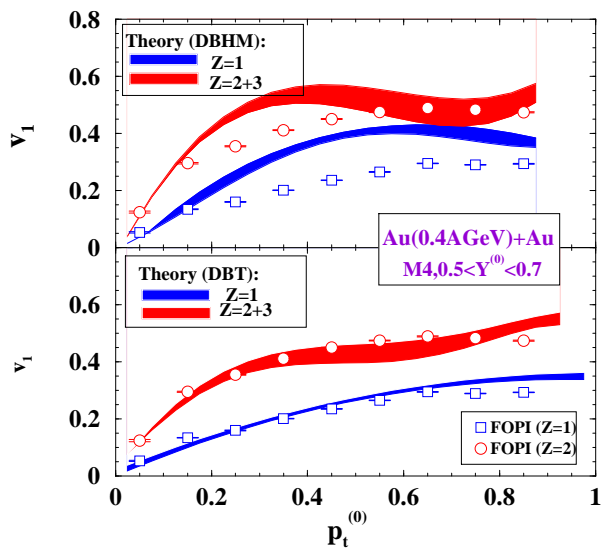}
}
\caption{In-plane flow in terms of the first Fourier coefficient 
(see Eq. (\protect\ref{fit})) for semi-central (PM4) Au+Au collisions at 0.4 AGeV for 
protons and light fragments (data are from \protect\cite{anton}). 
Statistical uncertainties of the calculations are indicated by bands.}
\label{Fig1}   
\end{figure}
A significant dependence of the differential in-plane flow on the DB model 
is observed. At low transverse momenta ($p_{t}^{(0)} \leq 0.2$) the 
two models do not show large differences. Since the 
mean transverse flow is dominated by low $p_{t}$ particles the $<p_x /A>$ observable 
therefore does not differentiate very much 
between the two models in the present situation. 
Consequently, in Fig.\ref{Fig5} the two curves for $<p_x /A>$ (0.4 AGeV) 
are very similar to each other in the considered rapidity range. 
However, the difference between the models becomes more pronounced 
when higher transverse momenta ($p_{t}^{(0)}\geq 0.2$) are studied. 
In Fig.\ref{Fig1} DBT reproduces the data very well for both, protons and 
light fragments, whereas DBHM overpredicts the 
flow significantly which reflects the 
its more repulsive character at baryon densities above $\rho_0$. 
The comparison to data favours a weaker repulsion at higher densities 
as predicted by DBT. In contrast to the global $P_{x}^{dir}$ 
observable the quantity $v_{1}(p_{t}^{(0)},~Y^{(0)})$ is more 
sensitive to high density matter in the participant region.

The situation is similar for light fragments ($Z=2+3$) which, 
however, show a stronger collectivity compared to free nucleons, 
as also seen by the higher value of $v_2$ in Fig.\ref{Fig1}. 
This can be understood by assuming that the heavier fragments are 
mainly formed in the spectator regions, whereas the free nucleons 
originate from the entire phase space. Thus, one obtains on the 
average a higher mean transverse momentum $<p_{x}/A>$ for fragments 
than for nucleons. This scenario of the fragment formation is consistent with the 
findings in Ref. \cite{gait00} that the spectator enters into an instability 
region with conditions which are near the experimental ones for 
a liquid-gas phase transition \cite{aladin00}. 

\subsection{Out-of-plane flows}
A preferential out-of-plane emission of particles, 
the so-called squeeze-out, is characterised 
by negative values of the second Fourier coefficient $v_{2}$ in 
Eq. (\ref{fit}) whereas a positive value of $v_2$ indicates in-plane flow 
\cite{povo98}. A related variable to characterise 
the azimuthal anisotropy of particle emission is the squeeze-out 
ratio $R_{N}$ defined by \cite{crochet97}
\begin{eqnarray}
R_{N}&=&(N(\Phi=90^{o})+N(\Phi=270^{o}))/(N(\Phi=0^{o})
\nonumber\\
&+& N(\Phi=180^{o}))=
(1-2v_{2})/(1+2v_{2})~~.
\end{eqnarray}
In terms of $R_{N}$ an isotropic emission corresponds to 
$R_{N}=1$, $R_{N} > 1$ indicates 
squeeze-out and  $R_{N} < 1$ a preferential in-plane emission. 
At lower energies squeeze-out is mainly due to shadowing of the participant 
particles by the spectators and it is therefore most pronounced for 
mid-rapidity particles. A sensitive way to probe the momentum dependence 
of the mean field is the transverse momentum dependence of the 
elliptic flow $v_{2}$. In contrast to 
central reactions where the squeeze-out is influenced by both, the static 
and the momentum dependent part of the mean field, peripheral reactions 
allow to decouple the momentum dependence to some extent 
from the static part. In Ref. \cite{dani00} this has been 
demonstrated for $v_2$ at high $p_t$ in peripheral reactions and was also 
confirmed by the investigations of \cite{larionov00}. Since high 
$p_t$ particles almost exclusively stem from the high 
density phase they probe the momentum dependence of the optical 
potential at supranormal densities, as was seen also in Fig. 13.

\begin{figure}
\centering
\resizebox{0.45\textwidth}{!}{
  \includegraphics{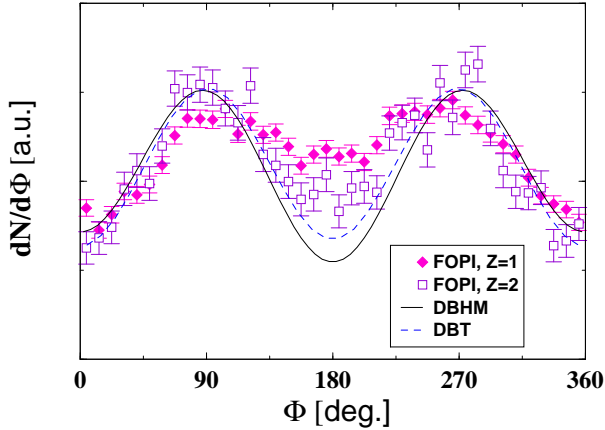}
}
\caption{Azimuthal distributions at mid-rapidity 
($|\Delta Y^{(0)}| \leq 0.15$) for semi-central Au+Au collisions 
at $0.6$ AGeV beam energy (data from \protect\cite{crochet2}).}
\label{Fig17}   
\end{figure}

First in Fig. \ref{Fig17} we consider semi-central (PM4) reactions. 
It shows the azimuthal distributions at mid-rapidity 
for all charged particles (nucleons plus fragments) in Au+Au at $0.6$ AGeV. 
The calculations are compared to experimental results from 
FOPI \cite{crochet2} for $Z=1$ and $Z=2$ fragments. 
The data show a stronger out-of-plane emission for $Z=2$ fragments 
relative to those for nucleons which is due to to the higher 
collectivity of fragments as discussed 
above. Due to limited statistics the theoretical values in Fig. \ref{Fig17} 
are given for all charged particles. We observe 
a qualitatively good agreement between theory and experiments, 
although both calculations slightly overestimate the experimental 
squeeze-out signal. 
\begin{figure}
\centering
\resizebox{0.45\textwidth}{!}{
  \includegraphics{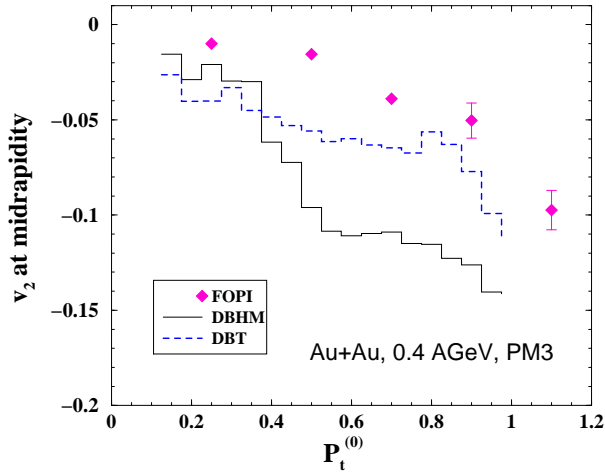}
}
\caption{Transverse momentum dependence of the elliptic flow at mid-rapidity 
($|\Delta Y^{(0)}| \leq 0.1$) 
$v_{2}$ for peripheral (PM3) Au+Au collisions at $0.4$ AGeV. Data are taken from 
\protect\cite{andronic99}.
}
\label{Fig2}   
\end{figure}

Fig.\ref{Fig2} shows the $p_t$ dependence of the 
elliptic flow $v_{2}$ in peripheral 
Au+Au collisions (PM3 multiplicity interval, $b_{\rm mean} \sim 7$fm) 
at 0.4 AGeV. 
Consistent with the picture that squeeze-out is mainly due to shadowing, 
which is most effective in the early high density phase of the 
reaction, the elliptic flow becomes increasingly negative 
with increasing $p_t$. A physical explanation for this behaviour is that 
a strong repulsive momentum dependent component of the nuclear interaction 
deflects these particles from the central zone more violently. 
This leads to enhanced shadowing by the spectator remnants inside the 
reaction plane and a stronger emission out-of-plane. This effect is 
clearly seen for the two types of interactions used here: Due to the stronger 
repulsive momentum dependence of the  DBHM forces at 
supranormal densities the squeeze-out signal increases 
much stronger with $p_t$ than for DBT, which on the other hand, is closer to 
the data of \cite{andronic99}. 

This behaviour is also consistent 
with the findings of Ref. \cite{andronic99} where soft and hard Skyrme forces 
within the framework of the QMD model 
were subjected to a comparison of the same observable. 
The softer equations-of-state (soft Skyrme in 
\cite{andronic99} and DBT in the present case) yield a slower 
increase of $v_2$ and are in reasonable agreement with 
experiment. However, in Ref. \cite{andronic99} both version of Skyrme forces 
have an identical momentum dependence whereas in the 
present case DBHM is significantly more repulsive at large densities. 
In the analysis of Ref. \cite{larionov00}, on the other 
hand, parameterisations were used which show the same density dependence, 
i.e. the same EOS, but differed in their momentum dependence. 
A stronger repulsive character of the model, expressed by both, 
a stiffer EOS and/or 
a stronger momentum dependence generally results in a stronger squeeze-out 
signal for high $p_t$ particles. The authors of Ref. \cite{larionov00} favour a 
parameterisation (HM in Ref. \cite{larionov00}) which 
yields results for this observable close to DBHM in our case. The 
comparison to the data of \cite{andronic99} in Fig.\ref{Fig2} 
supports, however, a weaker momentum dependence at 
supranormal densities.
\begin{figure}
\centering
\resizebox{0.45\textwidth}{!}{
  \includegraphics{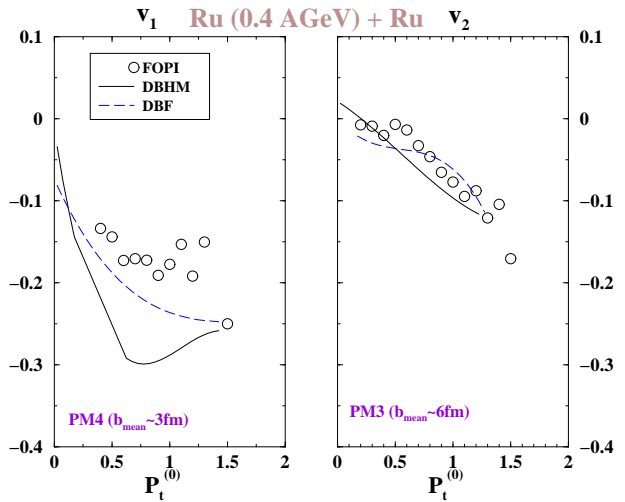}
}
\caption{
Transverse momentum dependence of the first ($v_{1}$, left) and 
second ($v_{2}$, right) azimuthal flow Fourier-coefficients for the system Ru+Ru 
at $0.4$ AGeV incident energy extracted in rapidity intervals of 
$-0.7 < Y^{(0)} < -0.5$ (left) and $-0.3 < Y^{(0)} < -0.1$ (right) 
for semi-central (PM4) 
and peripheral (PM3) reactions, respectively (data from \protect\cite{exp00}).}
\label{Fig12}   
\end{figure}
In Fig.\ref{Fig12} the same analysis is performed for the 
$Ru+Ru$ system. Here the differential components of the in-plane and 
out-of plane flow in terms of the Fourier-coefficients $v_{1}$ and $v_{2}$ 
as a function of the transverse momentum and rapidity are shown. The rapidity 
intervals are $-0.7 < Y^{(0)} < -0.5$ and $-0.3 < Y^{(0)} < -0.1$ for 
the $v_{1}$ and $v_{2}$ analysis, respectively. 
Both, DBHM and DBT are in 
fair agreement with the FOPI data \cite{exp00} on the $v_{2}$ observable. 
However, the $p_t$ dependence of the in-plane flow $v_1$ 
strongly depends on the model. As already observed 
for the $Au+Au$ system (Fig.\ref{Fig1}) DBHM strongly overpredicts 
the flow $v_1$ at intermediate $p_t$. Thus we conclude that 
the differences observed in the $v_1$ observable are due to the different 
momentum dependence of the fields. The $v_2$ observable shows here 
much less model dependence than e.g. in the $Au+Au$ system. 
In the smaller system the compressional effects are 
smaller which reduces the sensitivity to the 
different momentum dependence.

\begin{figure}
\centering
\resizebox{0.45\textwidth}{!}{
  \includegraphics{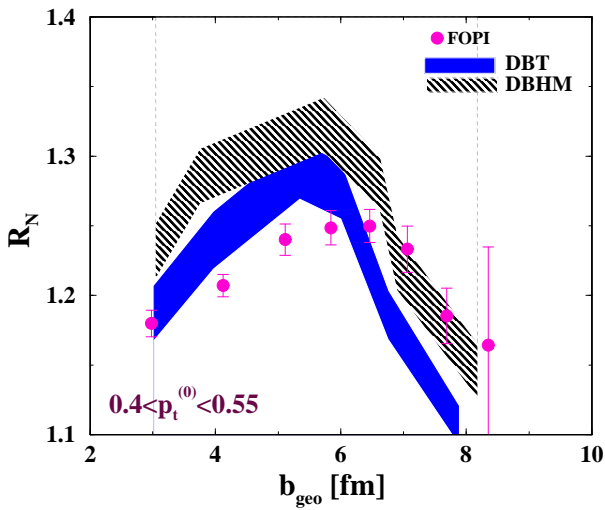}
}
\caption{Centrality dependence of the squeeze-out ratio 
$R_{N}$ at mid-rapidity ($|\Delta Y^{(0)}|<0.15$) for 
Au+Au collisions at $0.4$ AGeV (data from \protect\cite{crochet2}). 
Statistical errors of the calculations are indicated by bands.}
\label{Fig4}   
\end{figure}

In Fig. \ref{Fig4} the centrality dependence of the squeeze-out ratio $R_{N}$ at 
midrapidity ($|\Delta Y^{(0)}|<0.15$) 
for particles with high transverse momenta ($ 0.4 <p_{t}^{(0)}<0.55$) 
is shown. The decrease of the squeeze-out ratio $R_{N}$ at higher impact parameters 
can be explained by the strong vector repulsion of the nuclear mean field. 
Again both models can reproduce the centrality dependence of the 
experimental data qualitatively, but not in detail. The DBHM calculations overpredict 
the data at low impact parameters, whereas DBT underpredicts the data for peripheral 
collisions. Similar results have been found in recent studies of the FOPI collaboration 
in comparison with the IQMD model with soft/hard Skyrme forces \cite{exp00}. 

\begin{figure}
\centering
\resizebox{0.45\textwidth}{!}{
  \includegraphics{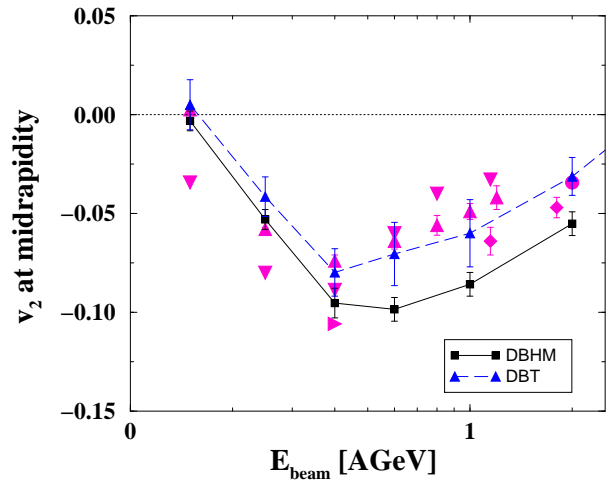}
}
\caption{Energy dependence of the elliptic flow $v_{2}$ at mid-rapidity. The data are 
taken from the FOPI- (triangles) \protect\cite{andronic99}, 
EOS (diamonds) \protect\cite{dani}, 
Plastic Ball (triangles down) \protect\cite{hermann}, LAND (triangle right) 
\protect\cite{hermann} and 
E895 (circle) \protect\cite{dani}.}
\label{Fig11}  
\quad . 
\end{figure}

Finally, Fig. \ref{Fig11} shows the excitation function of the total 
elliptic flow $v_2$ at mid-rapidity. With increasing 
beam energy the emission is preferentially perpendicular 
to the reaction plane ($v_{2}<0$) 
which approaches a maximum around $0.4$ AGeV and then decreases again. 
This behaviour can be understood by shadowing and compression 
effects as discussed in detail in Ref. \cite{dani}. 
Again both models reproduce the general trend of a sample of experimental 
data from FOPI 
\cite{andronic99}, EOS \cite{dani}, Plastic Ball \cite{hermann}, 
LAND \cite{hermann} and E895 \cite{dani} shown in  Fig.\ref{Fig11}. 
However, with DBHM the maximum of negative $v_2$ 
values is slightly shifted to higher energies and the absolute values are larger. 
The experimental data show strong variations in the 
magnitude of the $v_2$ coefficient which cover the range 
of the theoretical calculations. However, DBT better 
reproduces the maximum at the correct energy and lies closer 
in absolute magnitude to the 
most recent data from FOPI \cite{andronic99}. Assuming that the latter are 
the most reliable measurements, e.g. with respect to reaction plane 
corrections etc., the comparison to experiment favours the 
DBT mean fields.
\section{Summary}
We investigated the collective nucleon flow in heavy ion collisions 
at intermediate energies ($0.15\div 1$ AGeV) within a relativistic 
BUU transport model with mean fields based on relativistic 
Dirac-Brueckner-Hartree-Fock (DB) calculations for nuclear matter. 
The anisotropy of the local momentum space in the 
participant region of heavy ion reactions was taken into 
account in the colliding nuclear matter approximation.

We compared two different DB fields, those of ter Haar and 
Malfliet (DBHM) and the more recent calculations 
performed by the T\"ubingen group (DBT). From a theoretical point 
of view the latter ones have a stronger 
physical foundation since spurious contributions from a strong 
coupling of a pseudo-scalar $\pi NN$ vertex to negative energy states 
were removed by an improved projection scheme for the in-medium T-matrix. 
The DBT model EOS is slightly softer (K=250/230 MeV for DBHM/DBT) and the 
effective mass is significantly larger ($m^*=558/637$ MeV for DBHM/DBT) 
than for the DBHM model. This results in smaller fields 
and a less repulsive optical potential for DBT. 
The smaller repulsion of the DBT model is also expressed in a terms of a 
larger non-local effective mass ($m^{*}_{NR}=0.63/0.73~M$ for DBHM/DBT) 
which is commonly used in non-relativistic approaches in order to 
classify the strength of the momentum dependence of the potential. 

Both models yield a reasonable description of in-plane and out-of-plane 
flow observables. A more detailed comparison to data, in particular 
the transverse momentum dependence of $v_1$ and $v_2$ favours the 
softer EOS and the less repulsive character of the DBT optical 
potential at supranormal densities. This observation is consistent with 
the information obtained from subthreshold $K^+$ production where also 
the scenario of a soft EOS is supported \cite{sturm01,fuchs01}. 
At 0.6 and 0.8 AGeV DBT also yields a very accurate description 
of the transverse in-plane flow whereas at lower energies the in-plane flow 
requires some more repulsion as provided by the DBHM model. Interestingly, 
a similar observation was made in the non-relativistic approach of Ref. 
\cite{larionov00}. 

In summary, the microscopic DB approach, where no 
parameters are adjusted to the nuclear matter saturation properties nor 
to the empirical optical nucleon-nucleus potential, predicts a density 
and momentum dependence of the mean field which to a large extent is 
consistent with the observations from heavy ion collisions. 

\acknowledgments 
We are grateful to A. Andronic from FOPI for many discussions 
concerning the interpretation of data.


\begin{thebibliography}{}

\bibitem{buu}
E.A.~Uehling, G.E.~Uhlenbeck, Phys. Rev. {\bf 43} (1933) 552; 
P.~Danielewicz, Ann. Phys. {\bf 152} (1984) 239; 
G.F.~Bertsch, S.~Das~Gupta, Phys. Rep. {\bf 160} (1988) 189.

\bibitem{rbuu}
B.~Bl\"attel, V.~Koch, U.~Mosel, Rep. Prog. Phys. {\bf 56} (1993) 1.

\bibitem{hama}
S.~Hama, B.C.~Clark, E.D.~Cooper, H.S.~Sherif, R.L.~Mercer, 
Phys, Rev. {\bf C41} (1990) 2737; 
E.D.~Cooper, S.~Hama, B.C.~Clark, R.L.~Mercer, 
Phys. Rev. {\bf C47} (1993) 297.

\bibitem{skyrme}
J.~Aichelin, H.~St\"ocker, Phys. Lett. {\bf B176} (1986) 14.

\bibitem{ritter97}
W. Reisdorf and H.G. Ritter, 
Annu. Rev. Nucl. Part. Sci. {\bf 47} (1997) 663.

\bibitem{hermann}
N.~Herrmann, J.P.~Wessels, T.~Wienold, 
Annu. Rev. Nucl. Part. Sci. {\bf 49} (1999) 581, and refs. therein.

\bibitem{mbc94}
T. Maruyama, W.~Cassing, U.~Mosel, S. Teis and K.~Weber, 
Nucl. Phys. {\bf A573} (1994) 653.

\bibitem{zheng99}
Y.M. Zheng, C.M. Ko, Bao-An Li and Bin Zhang, 
Phys. Rev. Lett. {\bf A83} (1999) 2534.

\bibitem{nemeth99}
J. Nemeth, G. Papp, H. Feldmeier, Nucl. Phys. {\bf A647} (1999) 107.


\bibitem{giessen1}
P.K.~Sahu, A.~Hombach, W.~Cassing, M.~Effenberger, U.~Mosel, 
Nucl. Phys. {\bf A640} (1998) 493; 
A.~Hombach, W.~Cassing, S.~Teis, U.~Mosel, 
Eur. Phys. J. {\bf A5} (1999) 157.

\bibitem{giessen2}
P.K.~Sahu, W.~Cassing, U.~Mosel, A.~Ohnishi, 
Nucl. Phys. {\bf A672} (2000) 376.

\bibitem{dani00}
P. Danielewicz, Nucl. Phys. {\bf A673} (2000) 375.

\bibitem{larionov00}
A.B.~Larionov, W.~Cassing, C.~Greiner, U.~Mosel, Phys. Rev. 
{\bf C62} (2000) 064611.


\bibitem{gait96}
C.~Fuchs, T.~Gaitanos, H.H.~Wolter, Phys. Lett. {\bf B381} (1996) 23.

\bibitem{gait99}
T.~Gaitanos, C.~Fuchs, H.H.~Wolter, Nucl. Phys. {\bf A650} (1999) 97.

\bibitem{hs86}
        B. D. Serot, J. D. Walecka, 
        Advances in  Nuclear Physics, {\bf 16}, 1,
        eds. J. W. Negele, E. Vogt, (Plenum, N.Y., 1986)

\bibitem{ring96}
P. Ring, Prog. Part. Nucl. Phys. {\bf 78} (1996) 193. 

\bibitem{stars}
F. Weber, J. Phys. {\bf G25} (1999) R195.

\bibitem{IJMPE} \label{IJMPE}
B. D. Serot and J. D. Walecka, {\it Int.\ J.\ Mod.\ Phys.\/} 
E {\bf 6} (1997) 515.

\bibitem{FST} \label{FST}
R.J. Furnstahl, B.D. Serot, and H.-B. Tang,
{\it Nucl.\ Phys.\/} {\bf A615}, 441 (1997); 
H.W. Hammer, R.J. Furnstahl, 
{\it Nucl.\ Phys.\/} {\bf A678}, 272 (2000).

\bibitem{lutz00}
M. Lutz, B. Friman, Ch. Appel, Phys. Lett. {\bf B474} (2000) 7.

\bibitem{DBHF}
G.J.~Horowitz, B.D.~Serot, Phys. Lett. {\bf B137} (1984) 287; 
Nucl. Phys. {\bf A464} (1987) 613.

\bibitem{btm90}
   W. Botermans, R. Malfliet, 
   Phys. Reports {\bf 198} (1990) 115.

\bibitem{harmal}
B.~ter~Haar, R.~Malfliet, Phys. Reports {\bf 149} (1987) 207.

\bibitem{tueb2}
C.~Fuchs, T.~Waindzoch, A.~Faessler, D.S.~Kosov, 
Phys. Rev. {\bf C58} (1998) 2022.

\bibitem{tueb}
T.~Gross-Boelting, C.~Fuchs, A.~Faessler, 
Nucl. Phys. {\bf A648} (1999) 105.

\bibitem{DDH}
C.~Fuchs, H.~Lenske, H.H.~Wolter, 
Phys. Rev. {\bf C52} (1995) 3043.

\bibitem{bonn}
   R. Machleidt, {\em The Meson Theories of Nuclear Forces and Nuclear
   Structure\/},
   Advances in Nuclear Physics {\bf 19} (1989) 189.

\bibitem{machleidt99}
   R. Machleidt, nucl-th/9911059.

\bibitem{fuchs96}
C. Fuchs, E. Lehmann, L. Sehn, F. Scholz, J. Zipprich, T. Kubo, and 
Amand Faessler, Nucl. Phys. {\bf A603} (1996) 471.

\bibitem{exp00}
N. Bastid (FOPI collaboration), proceedings to 
\textit{Structure of the Nucleus at the Dawn of the Century}, 
(Publisher, Bologna 2000).

\bibitem{essler}
C.~Fuchs, P.~Essler, T.~Gaitanos, H.H.~Wolter, 
Nucl. Phys. {\bf A626} (1997) 987.

\bibitem{puri92}
R.K. Puri, N. Ohtsuka, E. Lehmann, A. Faessler, H.M. Martin, D.T. Khoa, 
G. Batko, and S.W. Huang, Nucl. Phys. {\bf A536} (1992) 201.

\bibitem{sehn}
L.~Sehn, H.H.~Wolter, Nucl. Phys. {\bf A601} (1996) 473.

\bibitem{gmat2}
D.T.~Khoa, N.~Ohtsuka, M.A.~Matin, A.~Faessler, S.W.~Huang, E.~Lehmann, 
R.K.~Puri, Nucl. Phys. {\bf A548} (1992) 102.

\bibitem{fuchs01}
C. Fuchs, Amand Faessler, E. Zabrodin, Y.M. Zheng, 
Phys. Rev. Lett. {\bf 86} (2001) 1974. 

\bibitem{gmat1}
   N. Ohtsuka, R. Linden,  A. Faessler, F.B. Malik,
   Nucl.Phys. {\bf A465} (1987) 550; 
   I. Izumoto, S. Krewald, A. Faessler, 
   Nucl.Phys. {\bf A341} (1980) 319;  

\bibitem{baldo89}
M. Baldo, I Bombaci, G. Giansiracusa, and U. Lombardo, 
Phys. Rev. {\bf C40} (1989) R491.  

\bibitem{zuo99}
W. Zuo, I. Bombaci, U. Lombardo, Phys. Rev. {\bf C60} (1999) 024605.  

\bibitem{lenske96}
F. de Jong and H.~Lenske, 
Phys. Rev. {\bf C54} (1996) 1488.

\bibitem{li93}
G.Q. Li and R. Machleidt, Phys. Rev. {\bf C48} (1993) 2707.

\bibitem{rqmd}
H.~Sorge, H.~St\"ocker, W.~Greiner, 
Ann. Phys. {\bf 192} (1989) 266.

\bibitem{Hu94}
S. Huber and J.Aichelin, Nucl. Phys. {\bf A573} (1994) 587.

\bibitem{hartnack98}
Ch. Hartnack, R.K. Puri, J. Aichelin, J. Konopka, S.A. Bass, 
H. St\"ocker, W. Greiner, Eur. Phys. J. {\bf A1} (1998) 151.

\bibitem{uma97}
V.S. Uma Maheswari, C. Fuchs, Amand Faessler, L. Sehn, 
D. Kosov, Z. Wang, Nucl. Phys. {\bf A628} (1998) 669.

\bibitem{fuchs95}
C.~Fuchs, H.H.~Wolter, Nucl. Phys. {\bf A589} (1995) 732.

\bibitem{andronic99}
A. Andronic {\it et al.} (FOPI collaboration), 
Nucl. Phys. {\bf A661} (1999) 333c.

\bibitem{dani}
C. Pinkenburg {\it et al. }, Phys. Rev. Lett. {\bf 83} (1999) 1295; 
P. Danielewicz, Roy A. Lacey, {\it et al. }, 
Phys. Rev. Lett. {\bf 81} (1998) 2438.

\bibitem{fopi95}
V.~Ramillien  {\it et al.} (FOPI--Collaboration), 
Nucl. Phys. {\bf A587} (1995) 802.


\bibitem{gait00}
T.~Gaitanos, H.H.~Wolter, C.~Fuchs, Phys. Lett. {\bf B478} (2000) 79.

\bibitem{crochet2}
P.~Crochet (FOPI--Collaboration), 
\textit{XXXIV International 
Winter Meeting of Nuclear Physics} (I. Iori, Bormio 1996); 
N.~Bastid {\it et al. } 
(FOPI--Collaboration), Nucl. Phys. {\bf A622} (1997) 573.

\bibitem{reisdorf97}
W.~Reisdorf, Nucl. Phys. {\bf A612} (1997) 493. 

\bibitem{crochet}
F. Rami {\it et al.} 
(FOPI Collaboration), Nucl. Phys. {\bf A646} (1999) 367.

\bibitem{zabrodin00}
E. Zabrodin, C. Fuchs, L.V. Bravina,. A. Faessler, 
Phys. Rev. {\bf C63} (2001) 034902.

\bibitem{povo98} A.~M.~Poskanzer and S.~A.~Voloshin,
Phys. Rev. {\bf C58} 1671 (1998) 1671.

\bibitem{anton}
A.~Andronic {\it et al. } (FOPI--Collaboration),
Phys. Rev. {\bf C64} (2001) 041604.

\bibitem{crochet97}
P.~Crochet {\it et al. } (FOPI--Collaboration), 
Nucl. Phys. {\bf A624} (1997) 755.


\bibitem{aladin00}
T. Odeh {\it et al.} (ALADIN Collaboration), 
Phys. Rev. Lett. {\bf 84} (2000) 4557.

\bibitem{sturm01}
C. Sturm {\it et al.} (KaoS Collaboration), 
Phys. Rev. Lett. {\bf 86} (2001) 39. 

\bibitem{typel}
S. Typel, T. v. Chossy, W. Stocker, H.H. Wolter, 
proceedings to \textit{Int. Symp. Nucl. Structure Physics}, 
ed. R. Casten et al. (World Scientific, 2001).

\bibitem{dafin}
F. Daffin, K. Haglin, and W. Bauer, Phys. Rev. {\bf C54} (1996) 1375.

\end{thebibliography}
\end{document}